\documentstyle[12pt]{article}
\def\beq{\begin{equation}}
\def\eeq{\end{equation}}

\def\veps{\varepsilon}
\def\ap#1#2#3 {Ann. Phys. (NY) {\bf#1} (19#2) #3}
\def\apj#1#2#3 {Astrophys. J. {\bf#1} (19#2) #3}
\def\apjl#1#2#3 {Astrophys. J. Lett. {\bf#1} (19#2) #3}
\def\app#1#2#3 {Acta. Phys. Pol. {\bf#1} (19#2) #3}
\def\ar#1#2#3 {Ann. Rev. Nucl. Part. Sci. {\bf#1} (19#2) #3}
\def\cpc#1#2#3 {Computer Phys. Comm. {\bf#1} (19#2) #3}
\def\err#1#2#3 {{\it Erratum} {\bf#1} (19#2) #3}
\def\ib#1#2#3 {{\it ibid.} {\bf#1} (19#2) #3}
\def\jmp#1#2#3 {J. Math. Phys. {\bf#1} (19#2) #3}
\def\ijmp#1#2#3 {Int. J. Mod. Phys. {\bf#1} (19#2) #3}
\def\jetp#1#2#3 {JETP Lett. {\bf#1} (19#2) #3}
\def\jpg#1#2#3 {J. Phys. G. {\bf#1} (19#2) #3}
\def\mpl#1#2#3 {Mod. Phys. Lett. {\bf#1} (19#2) #3}
\def\nat#1#2#3 {Nature (London) {\bf#1} (19#2) #3}
\def\nc#1#2#3 {Nuovo Cim. {\bf#1} (19#2) #3}
\def\nim#1#2#3 {Nucl. Instr. Meth. {\bf#1} (19#2) #3}
\def\np#1#2#3 {Nucl. Phys. {\bf#1} (19#2) #3}
\def\pcps#1#2#3 {Proc. Cam. Phil. Soc. {\bf#1} (#2) #3}
\def\pl#1#2#3 {Phys. Lett. {\bf#1} (19#2) #3}
\def\prep#1#2#3 {Phys. Rep. {\bf#1} (19#2) #3}
\def\prev#1#2#3 {Phys. Rev. {\bf#1} (19#2) #3}
\def\prl#1#2#3 {Phys. Rev. Lett. {\bf#1} (19#2) #3}
\def\prs#1#2#3 {Proc. Roy. Soc. {\bf#1} (19#2) #3}
\def\ptp#1#2#3 {Prog. Th. Phys. {\bf#1} (19#2) #3}
\def\ps#1#2#3 {Physica Scripta {\bf#1} (19#2) #3}
\def\rmp#1#2#3 {Rev. Mod. Phys. {\bf#1} (19#2) #3}
\def\rpp#1#2#3 {Rep. Prog. Phys. {\bf#1} (19#2) #3}
\def\sjnp#1#2#3 {Sov. J. Nucl. Phys. {\bf#1} (19#2) #3}
\def\spj#1#2#3 {Sov. Phys. JEPT {\bf#1} (19#2) #3}
\def\spu#1#2#3 {Sov. Phys.-Usp. {\bf#1} (19#2) #3}
\def\zp#1#2#3 {Zeit. Phys. {\bf#1} (19#2) #3}
\begin{document}
\begin{titlepage}
\begin{center}
{\Large \bf Theoretical Physics Institute \\
University of Minnesota \\}  \end{center}
\vspace{0.3in}
\begin{flushright}
TPI-MINN-97/23-T \\
UMN-TH-1604-97 \\
August 1997
\end{flushright}
\vspace{0.4in}
\begin{center}
{\Large \bf On domain shapes and processes in supersymmetric theories
\\}
\vspace{0.2in}
{\bf M.B. Voloshin  \\ }
Theoretical Physics
Institute, University of Minnesota \\ Minneapolis, MN 55455 \\ and \\
Institute of Theoretical and Experimental Physics  \\
                         Moscow, 117259 \\
\vspace{0.3in}
\end{center}
\begin{abstract}
A supersymmetric theory with several scalar superfields generically has
several domain wall type classical configurations which interpolate
between various supersymmetric vacua of the scalar fields. Depending on
the couplings, some of these configurations develop instability and
decay into multiple domain walls, others can form intersections in
space. These phenomena are considered here in a simplest, yet
non-trivial, model with two scalar superfields.
\end{abstract}
\end{titlepage}

\section{Introduction}
Domain walls in supersymmetric theories have attracted some attention in
recent literature both in a general theoretical aspect$^{\cite{cs}}$ and
in relation to a possible phenomenology of the early
Universe$^{\cite{morris}}$. Supersymmetric theories may posses a rich
non-trivial structure of domain walls. Indeed, in a renormalizable
4-dimensional theory with $N$ chiral superfields $\Phi_i$ ($i=1, \ldots,
N$) a generic superpotential $W(\Phi_i)$ is a cubic polynomial of the
fields. The supersymmetric vacua of the scalar fields $\phi_i$ are
determined by the equations $\partial W(\phi_i)/\partial \phi_i=0$.
These equations in general have $2^N$ solutions (with generally complex
fields $\phi_i$), each corresponding to a vacuum state $v_a$ ($a=1,
\ldots,\, 2^N$) with zero energy. Accordingly, there are $2^{N-1}
(2^N-1)$ domain wall configurations, each described by a solution to the
classical field equations that depends on only one coordinate ($z$) and
interpolates between different vacua $v_a$ and $v_b$ at two different
infinities in $z$, a configuration which here will be referred to as an
``$ab$ wall" or $w_{ab}$\footnote{It should of course be understood that
additional constraints, including the gauge symmetry constraints, do
restrict the form of the superpotential and can reduce the number of
possible vacua.} Depending on the couplings between the fields there can
be various relations between the energies of these
configurations. In particular, if  there is a vacuum state $v_c$ such
that the energy of the $ab$ wall, $\veps_{ab}$, is larger than the sum
of the energies of the walls $ac$ and $cb$: $\veps_{ac}+\veps_{cb}$,
then the $ab$ wall would decay into the pair $w_{ac}+w_{cb}$. In other
words instead of a well localized in $z$ transition between the domains
with the vacua $v_a$ and $v_b$ there arises an arbitrarily large domain
of the vacuum $v_c$, so that the localized transitions are $v_a \to v_c$
and $v_c \to v_b$. The latter picture can also be characterized as a
stratification of a two-phase configuration $(v_a,\, v_b)$ into a
three-phase one $(v_a,\,,v_c,\, v_b)$ This process immediately invites
the problem of whether the wall $w_{ab}$ is metastable or absolutely
unstable, i.e. whether there is or there is not an energy barrier
separating the configuration $w_{ab}$ from the $w_{ac}+w_{cb}$.

Another problem, related to the multitude of vacua in supersymmetric
theories, arises if one considers boundary conditions in the 3
dimensional space, involving more than one coordinate, namely with
different domains at different directions in space. Then the boundaries
between the domains have to intersect in space, thus suggesting the
problem of the stability of the intersections, and of the shapes formed
by the domain walls at the intersection. (It is known since long ago
that an intersection of domain walls in a theory of one scalar field is
not stable.)

A complete consideration of these problems looks rather complicated even
in a theory with two scalar fields (four vacua) if the superpotential is
assumed to be of a generic form of a cubic polynomial. In this paper
these problems are addressed in a recently considered$^{\cite{morris}}$
simplified case of a particular superpotential and partial results are
obtained illustrating the dependence of the stability of certain wall
configurations on the couplings in the model. The superpotential chosen
here for consideration is
\beq
W(\Phi,\, X)= \lambda \, X \, (\Phi^2 -a^2) + {1 \over 3} \, \mu X^3
\label{spot}
\eeq
with $\Phi, \, X$ being the superfields, $\lambda$ and $\mu$ being
dimensionless couplings, and, finally, $a$ being a dimensionful
parameter. The potential $V$ for the scalar components $\phi$ and $\chi$
of the superfields
\beq
V(\phi,\, \chi)=\left | \lambda \, (\phi^2-a^2) + \mu \, \chi^2 \right
|^2 + 4 \, \left | \lambda \, \phi \, \chi \right |^2
\label{pot}
\eeq
has four supersymmetric minima$^{\cite{morris}}$ : $\phi=\pm a, ~\chi=0$
and $\phi=0, ~\chi=\pm \sqrt{\lambda / \mu} \,a$. These vacuum states
are ascribed here numbers 1 through 4 as shown in Figure 1.
Correspondingly there are 6 domain wall configurations $w_{ab}$
connecting pairs of the vacua.

Due to the $Z_2 \times Z_2$ symmetry of the potential under reversing
the sign of either $\phi$ or $\chi$ the energies of the four domain
walls $w_{12}$, $w_{23}$, $w_{34}$ and $w_{41}$ are degenerate, while
the energies of the rest two wall configurations,the ``diagonals" (in
Fig.1) $w_{13}$ and $w_{24}$ are generally different. It is found here
that depending on the ratio of the coupling constants $\xi^2 =
\mu/\lambda\,$\footnote{It is sufficient to consider only the case of
positive $\mu/\lambda$, since the relative sign of the couplings can be
reversed by relabeling $\chi$ as $i \, \chi$.} at least one of the
latter
configurations can be unstable. Namely, for $\xi > 1$ the $w_{13}$
decays
into $w_{12}+w_{23}$ or $w_{14}+w_{43}$, while the ``diagonal" $w_{24}$
is stable at least locally, i.e. with respect to small perturbations at
all values of $\xi$. The ``diagonal" $w_{13}$ is stable locally for all
$\xi < 1$ and is also stable globally at small $1-\xi$. For $\xi=1$ the
energy of each ``diagonal" wall is
exactly equal to the sum of the energies of two ``sides", so that finer
effects, than considered here, determine the (in)stability of the
``diagonals". It is also found here that the instability of the $w_{13}$
at $\xi > 1$ is absolute in the sense that the
spectrum of small excitations around the  wall configuration develops a
runaway mode. Thus the
unstable wall is not a metastable local minimum of the energy in space
of field configurations interpolating between two corresponding
vacua\footnote{The results in this paper also illustrate that the local
instability of the ``diagonal" wall $w_{13}$ at  $\xi>1$ does not lead
to the so-called ribbon configurations$^{\cite{morris}}$ inside the
wall, but rather results in a complete decay of the wall into two stable
walls.} .
Needless to mention that the particular superpotential in
eq.(\ref{spot}) represents only a very limited class of possible models,
thus one can only hypothesize on the generality of this behavior.

The relations between the surface energies of the domain walls also
determine the possible shapes of intersections of the domain boundaries
in the cases where the boundary conditions at the space infinity require
presence of more than just two vacuum phases. We will show that the
equilibrium shapes of the intersections of the walls are determined by
the stratification considerations and by simple equations for the
intersection angles, analogous to those in the capillarity theory.

\section{Global stability and instability of  ``diagonal" walls}
Throughout this paper the parameters of the superpotential $\lambda,~
\mu$ and $a$ are assumed to be positive real. Then the vacuum states,
the field profiles, and also possible runaway modes are all described by
real fields $\phi$ and $\chi$.

The profile of the ``diagonal" walls (positioned across the $z$ axis and
centered at $z=0$) is given by the familiar one-field formulas:
\beq
w_{13}\,: ~~~~~~~~\phi(z)=a \,\tanh (\lambda \, a \, z)~, ~~~~
\chi(z)=0~;
\label{w13}
\eeq
\beq
w_{24}\,: ~~~~~~~~\chi(z)={1 \over \xi} \, a \, \tanh (\xi \, \lambda \,
a \, z)~, ~~~~ \phi(z)=0~.
\label{w24}
\eeq
The surface energy densities of these walls are respectively
$\veps_{13}= {8 \over 3} \, \lambda \, a^3$ and $\veps_{24}= {8 \over 3}
\, \lambda \, a^3 /\xi$.

The profiles of all four ``side" walls are related by the $Z_2 \times
Z_2$ symmetry of the model considered here and they have same energy.
Thus it would be sufficient to find the profile and the energy density
of one of these walls, e.g. $w_{12}$. However, an analytical solution is
not readily available, except for the case of $\xi=1$, where an
additional symmetry arises with respect to the permutation $\phi
\leftrightarrow \chi$. In this case the problem is reduced to the
one-field domain wall, by a $\pi/4$ rotation in the field space, and the
solution is
\beq
w_{12} \, (\xi=1)\,: ~~~~~~~~\phi={a \over 2} \left [ 1-\tanh (\lambda
\, a \, z) \right ]~, ~~~~ \chi={a \over 2} \left [ 1+\tanh (\lambda \,
a \, z) \right ]~.
\label{w12}
\eeq
The energy density of this configuration is $\veps_{12} (\xi=1)= {4
\over 3} \, \lambda \, a^3$, thus a two ``side" wall configuration (e.g.
$w_{12}+w_{23}$) has the same energy as a ``diagonal" wall ($w_{13}$) in
the limit where the two side walls are separated by large (formally
infinite) distance, so that their interaction is neglected.

In order to assess the relation between the sum of the energies of the
``side" walls and of the ``diagonal" one for $\xi \neq 1$ two approaches
are used here: perturbation over the configuration in eq.(\ref{w12}) for
small $|\xi -1|$ and a variational bound for the energy of the ``side"
walls at arbitrary $\xi$.

In order to apply the perturbation theory we write the potential in the
sector of real fields as a function of the fields $\phi$ and $\chi$ and
of the parameters $\lambda$ and $\xi$:
\beq
V(\phi,\, \chi, \, \lambda, \, \xi)=\lambda^2 \, \left [ (\phi^2+\xi^2
\, \chi^2-a^2)^2+ 4 \, \phi^2 \, \chi^2 \right ]
\eeq
and notice that from dimensional considerations the energy density of a
wall depends on the parameters as $\veps=\lambda \, a^3 \, f(\xi)$ with
the non-trivial information contained in the dimensionless function
$f(\xi)$. Thus one can apply
the standard perturbation theory formula for the dependence of the
energy density $\veps_{12}$ on $\xi$ at $\xi=1$ and find:
\beq
{d \veps_{12} \over d \xi}= \int \, \left. {\partial V(\phi,\, \chi; \,
\lambda, \, \xi) \over \partial \xi} \right |_{\phi=\phi_0,\,
\chi=\chi_0,\, \xi=1} \, dz=-{4 \over 3}\, \lambda  \, a^3~,
\label{pert}
\eeq
with $\phi_0$ and $\chi_0$ given by the exact solution (\ref{w12}) at
$\xi=1$. Comparing the energy of the ``side" walls in this order in
$|\xi-1|$ with the energy of the ``diagonal" walls described by the
equations (\ref{w13}) and (\ref{w24}), one finds that the wall $w_{13}$
is stable with respect to decay into two ``side" walls at $\xi < 1$ and
is unstable at $\xi > 1$. Naturally, this conclusion is valid only in a
finite region of $\xi$ near $\xi=1$. It will be shown by a variational
bound that the instability of the $w_{13}$ holds for all $\xi$ greater
than one. In the whole domain $\xi <1$, strictly speaking, only the
local stability of the wall $w_{13}$ will be established here. However
it would be quite surprising if there is a sub-domain of $\xi$ at $\xi
<1$ where the global stability of the wall $w_{13}$ is broken.
As to the ``diagonal" wall $w_{24}$, its energy coincides with the sum
of the energies of two ``side" walls in this order in $|\xi-1|$ and no
conclusion about its global stability can be made in this order. We
shall see however that the wall $w_{24}$ is locally stable for all
values of $\xi$ at least up to loop effects.

In order to obtain a variational bound for the energy density of the
``side" wall $w_{12}$ a `good' trial configuration should be constructed
interpolating between the vacua $v_1$ and $v_2$. We use here the obvious
property of the potential $V$ that it retains its quartic form under a
linear transformation of the fields. Thus the potential has a quartic
dependence along the straight line in the field space connecting the
vacua $v_1$ and $v_2$ in which vacua the potential along the line has
degenerate minima. Therefore if the trajectory of the trial
configuration in the field space is restricted to this line, the minimal
energy is given by the standard solution for the one-field wall. This
solution for the trial configuration is readily found as
\beq
\phi_t(z)={a \over 2} \left [ 1-\tanh (\lambda \, a \, z) \right ]~,
~~~~ \chi_t(z)={a \over 2 \, \xi} \left [ 1+\tanh (\lambda \, a \, z)
\right ]~.
\label{tsol}
\eeq
The energy of this configuration is given by $\veps_t={2 \over 3}  \,
(1+\xi^{-2}) \, \lambda \, a^3$, which is our upper bound for the actual
value of the energy of any of the ``side" walls. One can notice that
this upper bound is quite `good' in the sense that not only it
reproduces the actual energy at $\xi=1$
(as it should by construction) but also reproduces the slope of the
energy at $\xi=1$.

One can now see that at $\xi > 1$ the energy of the ``diagonal" wall
$w_{13}$ always exceeds $2 \, \veps_t$. Thus the wall $w_{13}$ is
unstable with respect to decay into a pair of ``side" walls. As to the
energy of the other ``diagonal", $w_{24}$, it is not greater than $2 \,
\veps_t$ at any $\xi$, and this trial configuration is of little use for
determining the stability of the $w_{24}$\footnote{As well as the whole
variational approach is not very helpful here, since the wall $w_{24}$
is most likely globally stable at all $\xi$.}.

\section{Local stability consideration}
The stability of the walls with respect to small perturbations is
analyzed by the standard method of linearizing the field equations near
the classical solution and finding the spectrum of the eigenvalues for
$\omega^2$ with $\omega$ being the frequency of the mode:
$\psi(t,x,y,z)=e^{-i\omega t} \, u(x,y,z)$. Negative eigenvalues for
$\omega^2$ correspond to runaway modes and imply the instability of the
configuration around which the mode expansion is performed. Conversely,
absence of negative modes means that the classical background is stable
at least locally, i.e. with respect to small perturbations.

The linearized equations have the form of the Schr\"odinger equation in
which the classical background provides the potential that depends on
$z$ under our convention about the placement of the walls. Negative
modes arise when the potential is sufficiently negative over a
sufficiently wide region of $z$. With real coupling constants and in a
real background, the energy of the modes in the imaginary direction of
the fields is always positive definite with the potential given by
eq.(\ref{pot}). Thus in a search for negative modes one can disregard
the perturbation of the fields in the imaginary direction and consider
only real fields.

The complete field equations in the model considered have the form
\beq
\partial^2 \, \phi + 2 \, \lambda \, \phi \left [ \lambda \, (\phi^2
-a^2) + \mu \, \chi^2 \right]+ 4 \, \lambda^2 \, \phi \, \chi^2=0~,
\label{peq}
\eeq
\beq
\partial^2 \, \chi + 2 \, \mu \, \chi \left [ \lambda \, (\phi^2 -a^2) +
\mu \, \chi^2 \right]+ 4 \, \lambda^2 \, \chi \, \phi^2=0~,
\label{ceq}
\eeq
where $\partial^2 = \partial_t^2 - \nabla^2$. Let us first consider the
perturbations over the wall $w_{13}$ for which the classical solution is
given by eq.(\ref{w13}). Since in this background $\chi=0$, the
linearized equations for the $\phi$-modes and the $\chi$-modes decouple:
the former modes are described by linearization of the equation
(\ref{peq}) in perturbation of $\phi$ after $\chi$ is set identically
equal to zero, while the latter modes are described by the linear in
$\chi$ part of the equation (\ref{ceq}) with the $\phi$ replaced by the
classical background. Thus the problem for the $\phi$-modes reduces to
the standard one-field situation in which it is known since long
ago$^{\cite{polyakov,mv}}$ that the lowest mode is the zero one and
there are no negative modes. Therefore a negative mode can arise only
from the linearized equation (\ref{ceq}) for the $\chi$-modes:
\beq
\partial^2 \, \chi + \lambda^2 \, a^2 \left ( 4 - {{4 + 2\, \xi^2} \over
\cosh^2 \, \lambda \, a \, z} \right ) \, \chi =0~.
\label{lceq}
\eeq
The lowest eigenvalue in this well known Quantum Mechanical problem is
given by
\beq
\omega_{min}^2= \lambda^2 \, a^2 \, \left [ 4 - {1 \over 4} \, \left (
\sqrt{17+8 \, \xi^2} -1 \right )^2 \right ]~.
\label{leig}
\eeq
One can readily see from this expression that $\omega_{min}^2$ becomes
negative for $\xi >1$. Thus the wall $w_{13}$ is unstable locally at
$\xi > 1$ and is stable under small perturbations at $\xi<1$.

For the other ``diagonal" wall, $w_{24}$, described by eq.(\ref{w24})
the meaning of the $\phi$ and $\chi$ modes is reversed: the linearized
equation (\ref{ceq}) describes the $\chi$ modes in a situation
equivalent to the one-field problem, while the linearized equation
(\ref{peq}), describing the $\phi$ modes should be inspected for a
possible presence of negative eigenvalues. The linearized equation for
the $\phi$ modes has the form
\beq
\partial^2 \, \phi + \xi^2 \, \lambda^2 \, a^2 \, \left ( {4 \over
\xi^4} - {{4+2\, \xi^2} \over \xi^4 \, \cosh^2 \, \xi \, \lambda \, a \,
z} \right ) \, \phi =0~.
\label{lpeq}
\eeq
The lowest eigenvalue in this equation is $\omega_{min}^2=0$ for all
$\xi$. Thus there are no runaway modes for the wall $w_{24}$ at least in
this approximation, and the wall is locally stable at all $\xi$, unless
radiative corrections push the $\omega_{min}^2$ into the negative
region.

To summarize the discussion of this section: with respect to small
perturbations the ``diagonal" wall $w_{13}$ is stable at $\xi<1$ and is
unstable at $\xi > 1$, while the other ``diagonal" $w_{24}$ is stable
for all $\xi$ at least in the approximation considered here.

\section{Shapes of the domains: intersections of walls}

The relations between the surface energies of the walls determine the
shapes of adjacent domains, containing different vacua. One simple
example of this has already been mentioned: if the wall $w_{13}$ is
unstable ($\xi > 1$), the domains with $v_1$ and $v_3$ cannot be
adjacent. There necessarily should be a domain with either $v_2$ or
$v_4$ separating them. Naturally, in the three dimensional space there
arises a whole variety of configurations involving more than two
different domains.

Let us first discuss the planar configurations, i.e. those where the
fields depend on two spatial variables. In a one-field theory it is
known that an intersection of domain walls is unstable. Indeed the
configuration shown in Fig. 2a has translational zero modes, whose shape
is given by the gradient of the field. Since there are directions in the
plane, where the field approaches the same values at both infinities,
the component of the gradient in such direction necessarily has zero.
Thus the zero mode cannot be the lowest in the spectrum and  a negative
mode exists, leading to a separation of the walls shown in Fig. 2b.
However if, as in supersymmetric theories, there are more than two
degenerate vacua, one can easily construct a ``triple intersection", as
shown in Fig.3. If the energy densities of the three domain walls
satisfy the triangle condition, i.e. each of the energies is less than
the sum of the other two, one can find the stable shape of such triple
intersection. Indeed, the wall acts as a film with the surface tension
given by the energy density $\veps$. Thus the relative angles between
the walls in the equilibrium shape are determined by the balance of
forces. In terms of the notation in Fig. 3 the conditions for this
balance are: $\veps_{ac} \, \sin \alpha = \veps_{bc} \, \sin \beta$ and
$\veps_{ac} \, \cos \alpha + \veps_{bc} \, \cos \beta= \veps_ab$. Viewed
as equations for the angles $\alpha$ and $\beta$ these conditions always
have a real solution, provided that the three surface tensions satisfy
the triangle condition. If the triangle condition is not satisfied, so
that e.g. $\veps_{ab} > \veps_{ac} + \veps_{bc}$ the equilibrium is
impossible and the configuration stratifies into one where the domain
$v_c$ separates the domains $v_a$ and $v_b$\footnote{Note, that in this
situation the equilibrium equations would have a solution with imaginary
$\alpha$ and $\beta$, which would correspond to the rapidities of the
walls $w_{ac}$ and $w_{bc}$ in the decay of the wall $w_{ab}$.}.

One can consider, as a logical possibility, an instability of the triple
intersection with respect to the process, shown in Fig.4, i.e. when a
fourth phase $v_d$ intervenes in the middle, providing a lower energy
due to a lower wall tension between the domain $v_d$ and each of the
three other domains. However a simple geometric consideration reveals
that the energy is lowered only in the case where one of the walls in
Fig. 3 is unstable with respect to decay into two walls with the domain
of $v_d$ in between. In this case the configuration of Fig. 3 is
unphysical since the domain with $v_d$ will intervene anyway.

Consider now a quadruple intersection as shown in Fig. 5a with each of
the four walls assumed to be stable. From a simple geometrical counting
of the total surface energy one concludes that such intersection should
split into two triple intersections, i.e. into the configuration of
either Fig. 5b or Fig. 5c, depending on which of the walls $w_{ac}$ or
$w_{bd}$ has lower energy and is stable. In the model considered in this
paper at least one of the latter walls is stable. It is not known at
present, whether this is valid in general case. Therefore one may
speculate that it may be that both the $w_{ac}$ and $w_{bd}$ are
unstable, and the shapes in both Fig. 5b and Fig. 5c do not exist in
equilibrium. Then the quadruple intersection of Fig. 5a has to be
stable. In this case the balance of tension forces leaves undetermined
one angle parameter. It is quite likely that this parameter is fixed by
finer local effects of the precise profiles of the fields at the point
of intersection.

In the two-field model considered here, the latter situation is not
realized and the quadruple, as well as higher, intersections split into
triple ones, which are the simplest non-trivial ones. The same is true
for the three dimensional intersections of the domain wall planes in
space. The minimal non-trivial intersection is a tetrahedral
intersection, where four domains meet at one point, and the relative
angles are determined by the balance of the surface tension forces. More
complicated intersection will split into ensembles of these simpler
ones, provided that the walls formed in the process of splitting are
stable.

\section{Summary}
The presented consideration of the structure of domain walls in a simple
two-field model with the superpotential (\ref{spot}) illustrates that
the phenomenology of the domain walls in supersymmetric models can be
quite feature-rich with a non-trivial dependence on the couplings. Some
of the walls can be unstable and decay into pairs of other walls at
certain values of the parameters. In the model considered here the
observed instability of a wall ($w_{13}$ at $\xi>1$) is complete, i.e.
the field of this configuration develops a runaway mode at the same
values of the appropriate coupling constant parameter at which the wall
is globally unstable. It is not known at present whether this is a
universal behavior or metastable walls can be constructed in more
general models. When regarded as candidates for a realistic model the
supersymmetric schemes with multiple domain structure can be tested
against the dynamics of the early Universe, where domain formation is
known to be subject to observational constraints$^{\cite{zko}}$. It
should be noticed however, that a consideration of the vacuum domain
structure in the early Universe may be complicated by thermal effects,
which in principle lift the energy degeneracy between at least some of
the vacua. Clearly, a better understanding of the phenomenology of the
domain walls both at $T=0$ as well as at a finite temperature would
require a much more extensive study than is presented in this paper.

\section{Acknowledgement}
This work is supported, in part, by the DOE grant
DE-AC02-83ER40105.\\[0.2in]

After this paper was finished, there appeared a paper by M.
Shifman$^{\cite{shifman}}$, where a degenerate class of domain walls in
supersymmetric models is considered as well wall stability and
instability for a superpotential equivalent to that in  eq.(\ref{spot})
discussed here.

\newpage
\unitlength 1mm
\thicklines
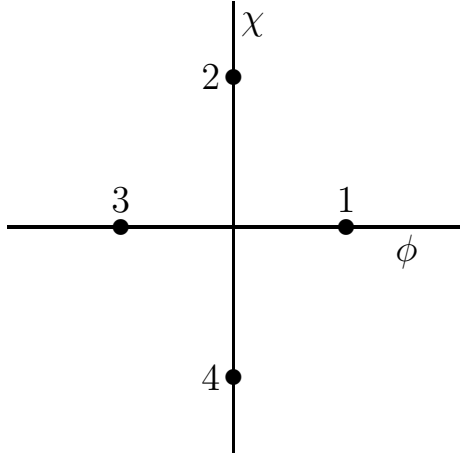
\begin{figure}
\begin{center}
\begin{picture}(70.00,70.00)
\put(10.00,40.00){\line(1,0){60.00}}
\put(40.00,10.00){\line(0,1){60.00}}
\put(25.00,40.00){\circle*{2.00}}
\put(55.00,40.00){\circle*{2.00}}
\put(40.00,60.00){\circle*{2.00}}
\put(40.00,20.00){\circle*{2.00}}
\put(41.00,67.00){\makebox(0,0)[lc]{{\large $\chi$}}}
\put(63.00,39.00){\makebox(0,0)[ct]{{\large $\phi$}}}
\put(55.00,42.00){\makebox(0,0)[cb]{{\large 1}}}
\put(37.00,60.00){\makebox(0,0)[cc]{{\large 2}}}
\put(25.00,42.00){\makebox(0,0)[cb]{{\large 3}}}
\put(37.00,20.00){\makebox(0,0)[cc]{{\large 4}}}
\end{picture}
\caption{Four supersymmetric vacua in a theory with the superpotential
(\ref{spot}).}
\end{center}
\end{figure}

\unitlength 1mm
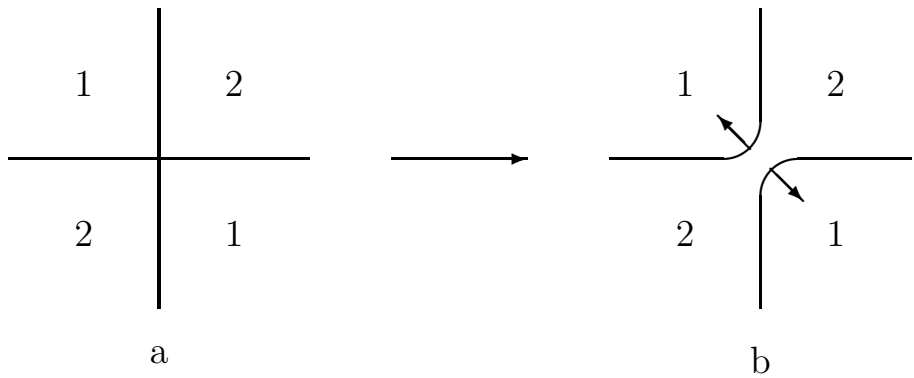
\begin{figure}
\begin{center}
\begin{picture}(125.00,45.00)
\put(5.00,25.00){\line(1,0){40.00}}
\put(25.00,5.00){\line(0,1){40.00}}
\put(85.00,25.00){\line(1,0){15.00}}
\put(110.00,25.00){\line(1,0){15.00}}
\put(105.00,45.00){\line(0,-1){15.00}}
\put(105.00,20.00){\line(0,-1){15.00}}
\put(110.00,20.00){\oval(10.00,10.00)[lt]}
\put(100.00,30.00){\oval(10.00,10.00)[rb]}
\put(15.00,35.00){\makebox(0,0)[cc]{{\large 1}}}
\put(35.00,35.00){\makebox(0,0)[cc]{{\large 2}}}
\put(35.00,15.00){\makebox(0,0)[cc]{{\large 1}}}
\put(15.00,15.00){\makebox(0,0)[cc]{{\large 2}}}
\put(95.00,35.00){\makebox(0,0)[cc]{{\large 1}}}
\put(115.00,35.00){\makebox(0,0)[cc]{{\large 2}}}
\put(115.00,15.00){\makebox(0,0)[cc]{{\large 1}}}
\put(95.00,15.00){\makebox(0,0)[cc]{{\large 2}}}
\put(106.33,23.67){\vector(1,-1){4.33}}
\put(103.67,26.33){\vector(-1,1){4.33}}
\put(25.00,0.00){\makebox(0,0)[ct]{{\large a}}}
\put(105.00,0.00){\makebox(0,0)[ct]{{\large b}}}
\put(56.00,25.00){\vector(1,0){18.00}}
\end{picture}
\caption{Intersection of domain walls in a one-field theory (a) is
unstable and decays into the configuration of the type (b) with the
walls moving in opposite directions.}
\end{center}
\end{figure}

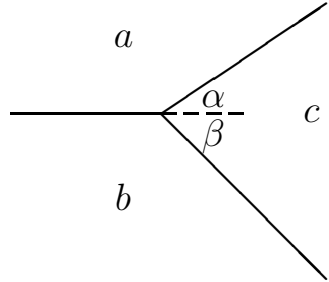
\begin{figure}
\begin{center}
\begin{picture}(52.00,39.67)
\put(10.00,25.00){\line(1,0){20.00}}
\put(30.00,25.00){\line(3,2){22.00}}
\put(30.00,25.00){\line(1,-1){22.00}}
\put(25.00,35.00){\makebox(0,0)[cc]{{\large $a$}}}
\put(25.00,14.00){\makebox(0,0)[cc]{{\large $b$}}}
\multiput(30.00,25.00)(3.00,0.00){4}{\line(1,0){2.00}}
\put(50.00,25.00){\makebox(0,0)[cc]{{\large $c$}}}
\put(37.00,27.00){\makebox(0,0)[cc]{{\large $\alpha$}}}
\put(37.00,22.00){\makebox(0,0)[cc]{{\large $\beta$}}}
\end{picture}
\caption{Intersection of three domain walls. The angles $\alpha$ and
$\beta$ are determined by the equillibrium of the tension forces.}
\end{center}
\end{figure}

\begin{figure}
\begin{center}
\begin{picture}(50.50,45.00)
\put(10.00,25.00){\line(1,0){20.00}}
\put(25.00,35.00){\makebox(0,0)[cc]{{\large $a$}}}
\put(25.00,14.00){\makebox(0,0)[cc]{{\large $b$}}}
\put(50.00,25.00){\makebox(0,0)[cc]{{\large $c$}}}
\put(30.00,25.00){\line(2,1){14.00}}
\put(44.00,32.00){\line(0,-1){16.00}}
\put(44.00,32.00){\line(1,2){6.50}}
\put(44.00,16.00){\line(1,-2){5.50}}
\put(39.00,25.00){\makebox(0,0)[cc]{{\large $d$}}}
\put(30.00,25.00){\line(3,-2){14.00}}
\end{picture}
\caption{A configuration, whose energy is lower than that of shown in
Fig. 3 only if one of the walls in Fig. 3 is unstable.}
\end{center}
\end{figure}
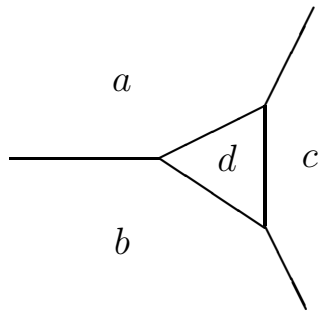

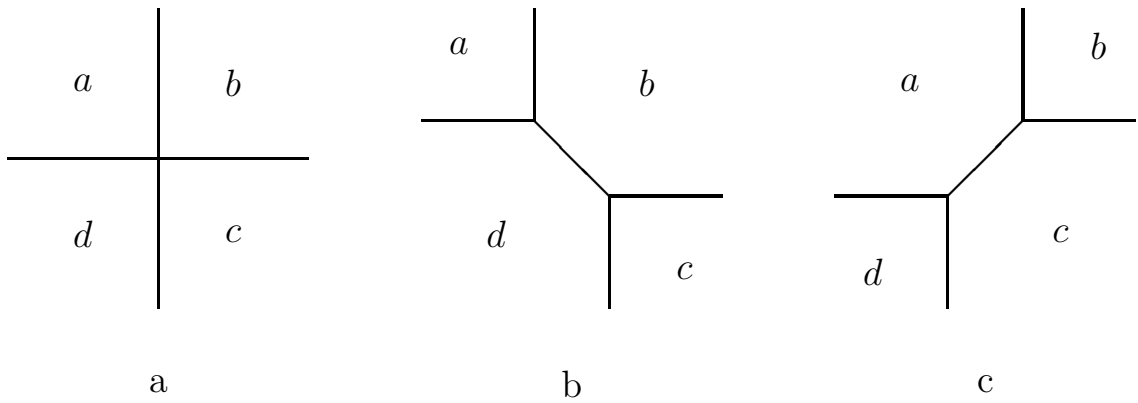
\begin{figure}
\begin{center}
\begin{picture}(155.00,50.00)
\put(5.00,30.00){\line(1,0){40.00}}
\put(25.00,10.00){\line(0,1){40.00}}
\put(60.00,35.00){\line(1,0){15.00}}
\put(75.00,35.00){\line(0,1){15.00}}
\put(75.00,35.00){\line(1,-1){10.00}}
\put(85.00,25.00){\line(0,-1){15.00}}
\put(85.00,25.00){\line(1,0){15.00}}
\put(115.00,25.00){\line(1,0){15.00}}
\put(130.00,25.00){\line(0,-1){15.00}}
\put(130.00,25.00){\line(1,1){10.00}}
\put(140.00,35.00){\line(0,1){15.00}}
\put(140.00,35.00){\line(1,0){15.00}}
\put(15.00,40.00){\makebox(0,0)[cc]{{\large $a$}}}
\put(35.00,40.00){\makebox(0,0)[cc]{{\large $b$}}}
\put(35.00,20.00){\makebox(0,0)[cc]{{\large $c$}}}
\put(15.00,20.00){\makebox(0,0)[cc]{{\large $d$}}}
\put(65.00,45.00){\makebox(0,0)[cc]{{\large $a$}}}
\put(95.00,15.00){\makebox(0,0)[cc]{{\large $c$}}}
\put(90.00,40.00){\makebox(0,0)[cc]{{\large $b$}}}
\put(70.00,20.00){\makebox(0,0)[cc]{{\large $d$}}}
\put(125.00,40.00){\makebox(0,0)[cc]{{\large $a$}}}
\put(150.00,45.00){\makebox(0,0)[cc]{{\large $b$}}}
\put(145.00,20.00){\makebox(0,0)[cc]{{\large $c$}}}
\put(120.00,15.00){\makebox(0,0)[cc]{{\large $d$}}}
\put(25.00,0.00){\makebox(0,0)[cc]{{\large a}}}
\put(80.00,0.00){\makebox(0,0)[cc]{{\large b}}}
\put(135.00,0.00){\makebox(0,0)[cc]{{\large c}}}
\end{picture}
\caption{An intersection of four domain walls (a) splits into two triple
intersections (the configuration (b) or (c)), provided that at least one
of the walls $w_{ac}$ and $w_{bd}$ is stable.}
\end{center}
\end{figure}

\end{document}